\documentstyle[preprint,aps]{revtex}
\begin{document}
\title{Effective-Medium Theory for the Normal State\\ in Orientationally
Disordered Fullerides}
\author{M.S. Deshpande, S.C. Erwin, S. Hong,
and E.J. Mele}
\address{Department of Physics and Laboratory for Research on
the Structure of Matter\\ University of Pennsylvania, Philadelphia,
Pennsylvania 19104}
\date{\today}
\maketitle
\begin{abstract}   An effective-medium theory for studying the electronic
structure of the orientationally disordered $M_3$C$_{60}$ fullerides
is developed and applied to study various normal-state properties.
The theory is based on a cluster-Bethe-lattice method in which the
disordered medium is modelled by a three-band Bethe lattice, into
which we embed a molecular cluster whose scattering properties are
treated exactly.  Various single-particle properties and the
frequency-dependent conductivity are calculated in this model, and
comparison is made with numerical calculations for disordered
lattices, and with experiment.  \end{abstract}


In the metallic fullerides $M_3$C$_{60}$ doped with the large alkali
metals, the fullerene molecules are quenched into an orientationally
disordered state \cite{steph}.  In this state, each of the C$_{60}$
molecules is centered on the sites of an fcc Bravais lattice and
adopts a high symmetry setting, with the twofold symmetry axes aligned
along [001] crystal directions.  There are two inequivalent ways of
achieving this orientation on each lattice site, and the intensities
of X-ray reflections measured for these solids are well described by a
merohedrally disordered structural model in which the choice of
setting varies randomly from site to site in the solid \cite{steph}.
It is now widely accepted that merohedral disorder is either a
dominant or contributing factor to a number of the observed electronic
properties of these doped systems \cite{hebard}.  In this paper we
apply an effective-medium theory to study the conduction-electron
states in these disordered solids.

The amplitude for electronic hopping across bond $\tau$ between
neighboring fullerene sites in these systems can be represented by a
real 3$\times$3 matrix, $T_{\mu\nu} (\tau)$, in which each of the
matrix elements corresponds to a possible choice of orbital
polarization $\mu(\nu) = x,y,z $ in the $t_{1u}$ manifold on the
initial and final sites
\cite{gelfa,gelfb,yild93,satpath}.  Orientational disorder on the fullerene
sites leads to a model with off-diagonal disorder in a Hamiltonian for
an electron which carries with it an internal ``orbital'' degree of
freedom, represented by a three component field.  Gelfand and Lu
\cite{gelfa} made the interesting observation that the sizes and signs of the
various matrix elements in $T$ can be changed by changing the relative
molecular orientations on the terminal sites.  This is a
representation-dependent statement, since the signs and sizes of the
various elements in $T$ are also changed by any local rotation of the
internal $t_{1u}$ orbital bases on either of the terminal sites.  Of
course, physically measurable quantities in this problem can always be
expressed in a representation-independent (gauge invariant) manner
that does not refer to a specific convention for defining the orbital
polarizations on the various molecular sites.  For example, the total
density of states is obtained from a trace over the single-particle
Green's function, $N(E) = -(1/\pi) {\rm Im} \, {\rm tr} \, G_+(E)$,
where the trace requires a sum over both sites and over the orbital
degrees of freedom on a single site.  If one considers a moment
expansion of the Green's function:
\begin{eqnarray}
G(E) &=& \left( \frac{1}{E} \right) + \left( \frac{1}{E} \right)
^2 H + \left( \frac{1}{E} \right) ^3 H H \nonumber \\   && + \left( \frac{1}{E}
\right) ^4 H H H
+ etc.\label{moment}
\end{eqnarray}
the product $H^n$ at $n$-th order in this expansion describes a possible
closed path of an electron containing $n$ steps in configuration space. The
required tr$\,H^n$ may depend on the molecular orientations encountered, but it
is independent of any particular orientation of the orbital basis on each site.

To study the product $H^n$, we examine a composite operator which
describes the forward and backward propagation of a particle along a
single bond $\tau$ in the manner
\begin{equation}
H_{bond}(\tau) = \left( \begin{array} {cc} 0&T_{M,N} \\
T^T_{M,N} & 0 \end{array} \right),\label{hbond}
\end{equation}
where $M$ and $N$ can each adopt the values ``A''or ``B'', denoting
the two possible orientational settings on the initial and final
sites. Diagonalizing this hopping operator, one sees that for any
hopping matrix $T$, the eigenvalues of the composite operator
$H_{bond}$ must be ordered in pairs $\pm\mid t_n \mid$, for which
their eigenvectors, $R^{\pm}_{n}$, are related by symmetry.  An
eigenvector describing a hop from site $i$ to site $f$ can be
constructed from two normalized three-component vectors $u_{\alpha} (
\alpha = 1,3)$, so that $R_n = (1/\sqrt{2})\left(u_{n,i},u_{n,f}
\right)$.  For each of these normalized vectors $u$, the three
components can be interpreted as three direction cosines, which locate
a point on the surface of a reference sphere in three dimensions and
define a set of principle axes (orbital polarizations) for the action
of the bond-hopping operator.

The elementary hop from $i$ to $j$ can therefore be represented by the
3$\times$3 matrix operation $H_{ji} = U^T_j h U_i$ where $h$ is a
diagonal matrix containing (by convention) only the three {\it
positive} eigenvalues of $H_{bond}$, and $U_i$ is a matrix whose rows
are the three components of the eigenvectors of $H_{bond}$ projected
onto the site $i$. Here the action of the internal hopping operator
$h$ is positive definite, and all sign conventions are then in the $U$
matrices, which rotate a state vector on the surface of the reference
sphere.  Interestingly, one finds that the diagonal elements of $h$ in
this problem depend only weakly on the molecular orientation!  As an
example, in Table I we list the diagonal elements of $h$ obtained from
the hopping matrices of \cite{gelfa}.  For the most important
(largest) of these terms, the fractional variation of the amplitude is
only of order 10 percent, so it is apparent that the dominant effects
of disorder in this problem are introduced through the polarizations
of the molecular orbitals, and not through the amplitude fluctuations.
We will therefore ignore this amplitude variation between the two
configurations and focus solely on the variation of the $U$ matrices
as we pass from bond to bond in the structure.  The $n$-th order terms
in Eq.~(\ref{moment}) then describe a process in which the internal
orbital polarization for a test particle rotates as it propagates on
its closed $n$-step trajectory on the lattice.

For propagation in a disordered medium in which the $U$'s are varying
randomly from bond to bond in the structure, for large loops one
expects a single reference orbital polarization to propagate
symmetrically over the reference sphere and not to contribute spectral
weight in the one-particle Green's function.  This cancellation rule
fails for an orientationally ordered crystal, since the large $n$-fold
loops remain highly correlated.  Finally, even for the orientationally
disordered system, at large order $n$ the special class of
``Brinkman-Rice'' \cite{brink} retraceable paths, which propagate
randomly from some reference site and exactly retrace their steps back
to the origin, always makes a positive-definite contribution to the
trace (even in this vector model) and determines the behavior of the
Green's function.

In view of this we now construct an effective medium for which {\it
only} the retracing paths are included in the single-particle Green's
function. This network has the topology of a tree, or Bethe lattice,
with one ``ingoing'' bond and eleven ``outgoing'' bonds at every node
of the network.  This network should be distinguished from the usual
simple Bethe lattice in that the electron field on each site is now
described by a three-component vector, and the hopping amplitudes
along each of the bonds are then given by the 3$\times$3 matrices
discussed above, which explicitly depend on the fcc bond orientations.
In the approximation that we ignore the weak amplitude fluctuations
which are listed in Table I, the intermolecular hopping matrices
between like (AA, BB) and unlike (AB, BA) molecular orientations can
be transformed into each other by suitable local rotations of the
orbital coordinate system on each lattice site. This local gauge
freedom is ultimately eliminated on a physical lattice by the closure
of rings of bonds in the network.  On the tree which contains no
closed rings we need consider only the dynamics in the fully
orientationally-ordered configuration.

The construction of the single particle Green's function for the
fulleride tree is now straightforward.  The Dyson expansion for the
Green's function $EG(E) = I + HG$ can be closed by introducing an
effective field $\Phi({\tau_{ij}};E)$ such that $G_{j,k} (E) =
\Phi({\tau_{ij}};E) G_{i,k} (E)$.  These $\Phi's$ are complex
energy-dependent 3$\times$3 matrices which satisfy the
self-consistency condition $ E \Phi(\tau) = H (\tau) + \sum _{\tau
\neq
\tau'} H
(\tau') \Phi (\tau') \Phi (\tau).
$ Once the $\Phi$'s are obtained, the diagonal block
of the one particle Green's function is  $ G_{0,0}  = [E - \sum
H (\tau) \Phi (\tau, E)]^{-1}$.

The density of states for the fcc fulleride tree is shown in Fig.~1.
The symmetry about $E=0$ follows from the fact that closed paths on
the tree require an even number of steps. The bandwidth obtained here
is very nearly the bandwidth obtained from simulations on lattices
with quenched orientational disorder, a feature which reflects the
fact that the retracing paths control analytic structure in the
Green's function for the disordered medium.

Interestingly, single-particle spectra obtained from numerical
simulations on lattice models with orientational disorder exhibit a
relatively large {\it asymmetry} with spectral weight enhanced at
negative energy in the conduction band.  This is a result of the phase
coherent propagation of an electron through the closed rings of bonds
on the fcc lattice, and the effect can be used to extract a refined
estimate of the elastic mean free path in the disordered system.  To
illustrate this we construct a series of clusters containing the
smallest closed threefold rings on the fcc lattice, and embed these
clusters within an effective medium treated in the tree approximation
developed above.  In the middle panel of Fig.~1 we show results
obtained for a tetrahedral prism in the fcc structure (four sites
which are mutually nearest neighbors), with the external bonds
tethered to limbs of the Bethe lattice.  Here, because of the presence
of closed rings of bonds, the orientational settings of the four sites
inside the cluster must be specified, and the figure shows results for
the three possible choices: A$_4$, A$_3$B, and A$_2$B$_2$. For the two
embedded tetrahedral clusters containing bonds between unlike
orientations, the spectra are asymmetric with spectral weight enhanced
at negative energies.

Similar calculations in which the size and symmetry of the embedded
cluster are varied show that the dephasing of the orbital polarization
introduced by the effective medium is a very strong effect in this
problem, due essentially to the large coordination number on the fcc
lattice.  Consequently, even for fully orientationally disordered
clusters, the effects of the short-range ring resonances through
several coordination shells must be retained before the data converge
suitably to those obtained from lattice simulations on completely
disordered fullerides.  In the bottom panel of Fig.~1, for example, we
display results obtained from embedded clusters containing randomly
chosen A$_{10}$B$_9$ and A$_{22}$B$_{21}$ molecular configurations.
These models just close the second (002) and third (112) coordination
shells, respectively, around a reference molecule at the origin.
After closing the third coordination shell on the lattice, the spectra
provide a relatively good description of numerical data (also shown)
obtained from a quenched average over an ensemble of orientationally
disordered 27-site lattices.  Apparently, in the orientationally
disordered solid the effective elastic mean free path is somewhat
larger than expected, and extends through several (typically 2--3)
coordination shells on the fcc lattice.  The physical point is that
one requires phase-coherent propagation of the electron over this
distance before the constructive interference from the short range
three-fold ring resonances can begin to compete with the very strong
dephasing of the electron orbital polarization introduced by its
coupling to the external medium.

The above discussion can now be extended to study the frequency dependence
of the conductivity at $T=0$,
\begin{equation}
\sigma(\omega) = \frac {e^2} {{2 \pi}
\omega} \int dE \, {\rm tr} \,
[ \hat{\jmath}^{\dagger} G(E) \hat{\jmath} G^{\dagger}
(E + \omega) ] \label{cond}
\end{equation}
where $G(E) = {\rm Re} \, G_+(E) \pm \, i \, {\rm sgn} (E-\mu) \, {\rm
Im} \, G_+(E)$, $\mu$ is the fermion chemical potential, and $
\hat{\jmath}_{ij} = (i/\hbar) H_{ij} \tau_{ij}$ is the current
operator in the $ij$-th bond. The integrand in Eq.~(\ref{cond})
describes a loop correlation function obtained from a trace over
products of Green's functions and vertex operators \cite{econ}.  The
conductivity calculated in this effective medium is most naturally
formulated in terms of a nonlocal bond-bond response function $\sigma
(r,r')$ so that $ j(r) = (1/\Omega_c) \int dr' \sigma (r,r') E(r') $.
Integration over all $r$ and $r'$ yields the $q=0$ limit of the
correlation function which is measured with optical probes.  Here, the
behavior of $\sigma (r,r')$ for large $\mid r - r' \mid$ can be
calculated exactly by introducing a vertex function, $\Gamma(\tau ;
E,E')$, which describes the contributions to the current density from
processes in which an electron and hole, created inside a reference
bond, propagate outward {\it along the same path} in the effective
medium, and ultimately recombine in an external bond of the network.
This vertex function plays an important role in this problem and can
be calculated from the self consistency condition \cite{ustbp}
\begin{eqnarray}
\Gamma({\tau};
E,E') &=& \Gamma^{(1)} ({\tau}; E,E') \nonumber \\&& + \sum _{\tau'}
\Phi({\tau}, E) \Gamma({\tau'};E,E')
\Phi ({\tau},E') \label{selfcon}
\end{eqnarray}
where $\Gamma ^{(1)} (\tau;E,E') =  \hat{\jmath}(\tau)
\Phi({\tau,E')} + \Phi(\tau,E)
\hat{\jmath}(-\tau) $.   The full current operator entering our discussion then
takes the form $\hat{\jmath}_{ij} = \hat{\jmath}^{(0)}_{ij} + \delta_{ij}
\Gamma_{i} (E,E')$,
where the first term is the bare current-density operator acting
inside a reference bond of our cluster, and the second term is a
(site-diagonal) boundary term which sums all the external
contributions to the current density due to excitations which leave
the cluster through that boundary site.

Our results for the zero-temperature conductivity as a function of
frequency on the fulleride tree are shown in Fig.~2.  The top panel
shows the separate contributions from the intra-bond diagonal part of
the conductivity $\sigma (r,r) $ (solid), and the $q=0$ contribution
from the inter-bond off-diagonal elements $(1 - \delta_{r,r'})
\sigma(r,r')$ summed over $r'$ (dashed).  We find that the intra-bond
contribution has a low-frequency value near 1750 S/cm.  By contrast,
the nonlocal piece provides a contribution of $\approx$ 650 S/cm at
zero frequency, and decreases monotonically with increasing
frequency.  This inter-bond term carries roughly one-fourth of the
total spectral weight in the system, and is the remnant in the
disordered model of the damped Drude response expected of a normal
metal. The total conductivity obtained by summing these contributions
is shown in the lower panel; it is very weakly frequency dependent in
the range $ 0 < \omega < 15t$, and has a limiting value $\approx$ 2400
S/cm at low frequency.

The separation of diagonal and off-diagonal components of the nonlocal
conductivity provides a very natural method for distinguishing the
inter- and intra-band contributions to the response function for this
highly disordered medium.  However, the two are more difficult to
separate solely on the basis of their frequency dependence.  In fact,
this is true of the experimental data as well \cite{rott,degeorg}, and
has led to some uncertainty as to the appropriate separation of the
conductivity into contributions from the ``bound'' charge and that
from the diffusive low-frequency dynamics of the ``free'' conduction
charge in the disordered medium.

The data presented in Fig.~2 describe the conductivity calculated
numerically for orientationally disordered superlattices \cite{gelfb}
reasonably well.  We observe that the residual resistivity calculated
for the tree is 25 percent larger than that reported from previous
theoretical work, due to the fact that the Bethe lattice yields a
systematic underestimate of the effective elastic mean free path in
this system. From our discussion above, the effective mean free path
is $\ell\approx$18 \AA, while for the bare tree this correlation
length is clearly smaller, of order the intermolecular spacing.  As a
test of this conjecture, we have replotted these data collapsing {\it
only} the inter-bond (dashed) contribution into a low frequency Drude
peak with a natural width of $\gamma \approx$ 6t, giving a good
description of the superlattice data out to $\omega \approx 15t$,
while slightly overestimating it for higher frequencies.

Experimentally, the conductivity shows a broad midinfrared peak
\cite{rott,degeorg}  near
1200 cm$^{-1}$, and comparison with the data of Fig.~2 then provides
an experimental estimate of the energy scale parameter, $t\approx 10$
meV. This value is reassuringly close to the value of $t=14$ meV
obtained from a comparison with the band calculations on the fully
ordered phase, using local-density theory \cite{erwin}.
Extrapolations of the conductivity obtained from an analysis of the
measured reflectivity spectra on K$_3$C$_{60}$ at 25 K yield $\sigma
(0) \approx$ 1400 S/cm \cite{degeorg}.  An independent low-temperature
extrapolation by the Berkeley group \cite{zettl} yields a slightly
larger value of $\sigma (0)
\approx$ 2000 S/cm,
in closer correspondence with the theory presented here.

We should mention that elastic scattering from the residual
``amplitude'' fluctuations in the electronic Hamiltonian, neglected in
our model, will provide an additional elastic scattering mechanism for
orientationally disordered solids, although we expect these to be
quite weak.  More importantly, the residual effects of
electron-electron interactions, which are generally regarded as
significant in the solids, and which are completely neglected in this
treatment, can be expected to systematically suppress the
low-frequency conductivity.  A quantitative theory for treating these
effects in this class of disordered conductors has yet to be
developed.

Acknowledgements: This work was supported by the NSF under the Grant  DMR
91-20668 and by the Department of Energy under Grant 91ER45118.

\begin{table} \caption{The three positive eigenvalues of the bond hopping
operator $H_{bond}$ for transitions between molecules with like (AA) and unlike
(AB) settings. The eigenvalues of $H_{bond}$ are ordered in pairs $\pm \mid h
\mid$.  } \begin{tabular}{cc} AA& AB  \\
\tableline
4.445 & 5.029  \\
1.910 & 2.670  \\
0.255 & 0.431  \\
\end{tabular}
\end{table}

\begin{figure}
\caption{Densities of states calculated for the fulleride tree and for
molecular clusters embedded in the tree. Here and elsewhere in the
paper the energy axis is in units of $t\approx 10$ meV, which sets the
energy scale for electronic hopping between neighboring sites.  (Top
panel) Density of states for the tree. (Middle panel) Densities of
states for three embedded tetrahedral clusters with various molecular
settings.  (Bottom panel) Densities of states at the core of embedded
19-site (dashed) and 43-site (heavy solid) clusters. The noisy curve
gives the results of a quenched average on an orientationally
disordered lattice. \label{fig1}}
\end{figure}

\begin{figure}
\caption{Frequency-dependent conductivity on the Bethe lattice. (Top) The
intra-bond (solid) and inter-bond (dashed) contributions to the
nonlocal conductivity.  (Bottom) The total conductivity (heavy curve)
is compared with results obtained from a disordered lattice model
containing 32 molecular sites (histogram).  In the light curve the
spectral weight in the inter-bond contribution is collapsed to a
lifetime-broadened Drude peak with a phenomenological width $\gamma$ =
6t.
\label
{fig2}}
\end{figure}

\end{document}